\documentclass{aa}
\usepackage{graphicx}
\usepackage{amsmath}
\usepackage{amssymb}
\usepackage{txfonts}
\usepackage{booktabs}
\usepackage[colorlinks=true,linkcolor=blue,allcolors=blue]{hyperref}
\usepackage[cmyk]{xcolor}
\usepackage{threeparttable}
\usepackage{placeins}

\newcommand{\Msun}{M$_\odot$}

\newcommand{\anglerv}{$\theta_{\textbf{r},\textbf{v}}$} 
\newcommand{\anglemo}{$\theta_{\textbf{L}_{\rm main},\textbf{L}_{\rm orbit}}$}
\newcommand{\angleso}{$\theta_{\textbf{L}_{\rm sat},\textbf{L}_{\rm orbit}}$}

\newcommand{\afanglemo}{$\langle \theta_{\textbf{L}_{\rm main},\textbf{L}_{\rm orbit}}' \rangle$}
\newcommand{\afangleso}{$\langle \theta_{\textbf{L}_{\rm sat},\textbf{L}_{\rm orbit}}' \rangle$}
\newcommand{\aanglerv}{$\langle \theta_{\textbf{r},\textbf{v}} \rangle$}

\newcommand{\aanglemo}{$\langle \theta_{\textbf{L}_{\rm main},\textbf{L}_{\rm orbit}} \rangle$}
\newcommand{\aangleso}{$\langle \theta_{\textbf{L}_{\rm sat},\textbf{L}_{\rm orbit}} \rangle$}

\newcommand{\HI}{{\sc H\,i }}

\begin{document} 

   \title{The effects of the orbital configurations of mergers on reshaping galaxy structures}

   \author{Xinyi Wu$^{1,2}$\thanks{xywu@pmo.ac.cn}, Ling Zhu$^3$\thanks{Corr Author: lzhu@shao.ac.cn}, Jiang Chang$^1$\thanks{changjiang@pmo.ac.cn}, Guangquan Zeng$^{4}$, Yu Lei$^{3,5}$
   }

   \institute{
   Purple Mountain Observatory, Chinese Academy of Sciences, Nanjing 210008, China
   \and
   School of Astronomy and Space Sciences, University of Science and Technology of China, Hefei 230026, China
   \and
Shanghai Astronomical Observatory, Chinese Academy of Sciences, 80 Nandan Road, Shanghai 200030, China
    \and
    Department of Physics, The Chinese University of Hong Kong, Shatin, N.T., Hong Kong S.A.R., China
    \and
    School of Astronomy and Space Science, University of Chinese Academy of Sciences, Beijing 100049, China
    \\
\email{lzhu@shao.ac.cn}
             }
   \date{Received; accepted}
   
   \titlerunning{The effects of mergers orbits}
\authorrunning{Wu et al.}  
 
  \abstract
   {
   We performed a systematic analysis of how the  orbital configuration of a merger can influence the structural formation of remnant galaxies using 531 merger pairs selected from IllustrisTNG-100.
   We comprehensively described the merger orbital configuration, considering the relative orbital motion of the merger pair and their disk orientations. We quantified the galaxy structures by dynamically defining four components: bulge, disk, warm component, and hot inner stellar halo.
   For mergers on spiral-in orbits, the disk planes of the two merging galaxies tend to be aligned with the orbital plane, leading to higher fractions for the disk and warm components, as well as lower fractions for the bulge and hot inner stellar halo components in the remnant galaxy. For mergers on direct collision orbits, the disk planes of the two galaxies tend to be perpendicular to the orbital plane, leading  to lower fractions for disk and warm components, as well as higher fractions of the bulge and hot inner stellar halo in the remnant. Mergers can lead to either an increase or decrease in the disk and bulge mass fraction in the remnant compared to the progenitor galaxy, depending on the merger orbital configurations; however, in 93\% of cases, mergers cause an increase in the hot inner stellar halo.  
   As a result, the luminosity fraction of the hot inner stellar halo (but not the bulge) in galaxies at $z=0$ is highly correlated with its total ex situ stellar mass.
   In addition, we find that merger on spiral-in orbits is one of the possible reasons for the formation of recently discovered red but \HI-rich (RR) galaxies.
   }


\maketitle

\section{Introduction}

In the framework of the Lambda cold dark matter (LCDM) universe, galaxies undergo hierarchical growth. Mergers play an important role in galaxy formation and evolution \citep{White1978,heckman1990nature,hopkins2008dissipation} and believed to be one major process leading to the quenching of galaxy star formation \citep{mihos1995gasdynamics,man2018star} and reshaping of galaxy structures \citep{2022MNRAS.513.1867D}. The diverse merger histories provide an explanation for the diverse galaxy morphology that we observe across the Hubble sequence \citep{Hubble1926}. 
The gas cooling and condensation result in gaseous disks with high angular momentum; most stars in the galaxy are thus formed in highly rotating disks in a quiescent environment. Merger is taken as a major physical process that destroys disks and leads to the formation of classic bulges and the growth of stellar halos \citep{cox2006kinematic,pillepich2015building,zhu2022mass}. 

However, galaxy mergers are complex processes. 
The morphology of the remnant galaxy is influenced by a combination of factors such as merger mass ratios, gas fractions, and merging orbits. Major mergers with similar masses of the two progenitor galaxies could severely destroy the disks of the progenitors and create an elliptical remnant \citep{toomre1977evolution,negroponte1983simulations,di2007star,hopkins2009effects,conselice2009structures,naab2014atlas3d,deeley2017galaxy}. On the other hand, minor mergers are typically less violent; they may not completely destroy the disks, but can still disturb the stellar disk of the main progenitor galaxy, leading to disk instability \citep{dekel2009formation,fiacconi2015argo,zolotov2015compaction,welker2016rise}. Stars accreted from minor mergers contribute significantly to the growth of the stellar halo and, thus, to the growth of the galaxy size, especially for elliptical galaxies \citep{bournaud2007multiple,naab2009minor,amorisco2017contributions}. 

The gas fraction also plays a crucial role in the process of galaxy mergers. The gas-poor major mergers are taken as an important formation path of the slowly rotating giant elliptic galaxy \citep{naab2006influence,hopkins2009effects}. For major mergers with a relatively high gas fraction, gas tends to flow towards the center of the galaxy, where the starburst occurs and a substantial amount of stars form, resulting in an increased bulge-to-total ratio (B/T) in the remnant; it could also lead to an elliptical morphology \citep{hopkins2008dissipation}. However, it has been found that when the gas content is even higher, the excess gas will gradually reform a stellar disk in the remnant galaxy in about 1-2 Gyr after the merger, resulting in a lenticular or even spiral galaxy \citep{1996ApJ...464..641M,springel2005formation,sotillo2022merger}.

In addition, the merger orbits and disk orientation of the progenitor galaxies also have an important influence on the morphology of the remnant galaxies. Disks are not always destroyed, even by major mergers; they survive in several cases. The disks of the main progenitors are found to survive largely when the satellite galaxy gradually spirals into the main progenitor, especially when the angle between the velocity vector and the position vector between the satellite and the main galaxy (also referred to as "collision angle") is close to 90 degrees \citep{zeng2021formation}. Furthermore, \cite{sotillo2022merger} pointed out that the combination of collision angle and mass ratio could help predict whether the disk will survive a merger or not. 
When the satellite galaxy has a large disk, the disk orientation of the satellite galaxy also plays a significant role. The disk of the main progenitor is more likely to be destroyed when the disk of the satellite is counter-rotating, as compared to the disk of the main galaxy \citep{hopkins2009disks,bois2011atlas3d,taylor2018origin,genel2015galactic,lagos2018quantifying}. 

The above works primarily focus on the survival of disk after the merger, lacking a systematic investigation into the formation of different structures in the remnant galaxies. Galaxy structures, such as the classic bulge and halo, are usually taken as indicators of the galaxy merger history. Traditionally, the classic bulge has been considered a key indicator of galaxy merger history; however, it is complicated by the diverse origin of bulges \citep{2024MNRAS.529.4565H, zhang2025diverse}. Instead, the outer faint stellar halo, which forms as a result of mergers, is widely used to quantify the total ex situ stellar mass of a galaxy \citep{2010ApJ...725.2312O,tacchella2019morphology,remus2022accreted}. Recently, the luminosity of the inner stellar halo has also been shown to be a good indicator of ex situ stellar mass \citep{davison2020eagle,zhu2022mass,zhu2022fornax3d}. However, there are still significant scatters in all these relations of structural parameters versus ex situ stellar mass.
The complication in mergers, like the merger ratio, gas fraction, and merger orbits, likely leaves an imprint on the formation of all different galaxy structures and might account for some scatters in these relationships indicating galaxy ex situ stellar mass. 

In this study, we systematically investigate the effects of merger orbital configurations on the formation of different structures in the remnant galaxies. We describe the merger orbital configurations in a comprehensive way, including the relative moving orbits of the merger pairs and the disk orientations of the merging galaxies. We select about 500 merger pairs with a wide range of mass ratios from TNG100. We dynamically decompose each galaxy into four components: a cold disk, a warm component, a hot and compact bulge, and a hot and extended halo following \citet{zhu2022mass}. We quantify the formation of all these components across these merger pairs and analyze their dependence on the merger orbital configurations; we provide an explanation for the relationship between the hot inner stellar halo and ex situ stellar mass in \citet{zhu2022mass}. 

This paper is organized as follows. In Section \ref{sec:methods}, we introduce the simulation, galaxy structure decomposition, the selection of galaxy merger pairs from the simulation, and the description of merger orbital configurations. In Section \ref{sec:results}, we present the results, including the influence of merger orbital configurations on disk survival, on formation of different galaxy structures, and an explanation for the correlation between the hot inner stellar halo and ex situ stellar mass. In Section \ref{sec:discussion}, we discuss the potential relationship between the merger orbit and the recently discovered red but \HI-rich(RR) galaxies. We summarize our findings in Section \ref{sec:summary}.

\section{Data and methods}
\label{sec:methods}

\subsection{TNG100 simulation}
\label{sec:TNG}

IllustrisTNG (\cite{springel2018first,marinacci2018first,naiman2018first,pillepich2018first,nelson2018first,nelson2019illustristng}, hereafter TNG), a suite of large cosmological hydrodynamical simulations and the successor of Illustris\citep{vogelsberger2014introducing,vogelsberger2014properties,genel2014introducing,nelson2015illustris}, is primarily aimed at helping to improve the understanding of the formation and evolution of galaxies, as well as to provide guidance for subsequent observations. 
The TNG simulations have been particularly successful in reproducing a wide range of observational findings \citep{nelson2019illustristng}. These include the galaxy mass-size relation at $0<z<2$ as mentioned in \cite{genel2018size}, as well as the sizes and heights of gaseous and stellar disk sizes and heights \citep{pillepich2019first}, galaxy colors, the stellar age, and metallicity trends at $z\sim0$ as a function of galaxy stellar mass in comparison to SDSS results \citep{nelson2018first}, resolved star formation in star-forming galaxies \citep{nelson2021spatially}, the characteristics of the stellar orbit distributions from the CALIFA survey \citep{xu2019study,zhang2025diverse}, and the kinematics of early-type galaxies in comparison to data from ATLAS-3D, MaNGA, and SAMI \citep{pulsoni2020stellar}. 

\begin{figure}[h!]
\centering\includegraphics[width=9cm]{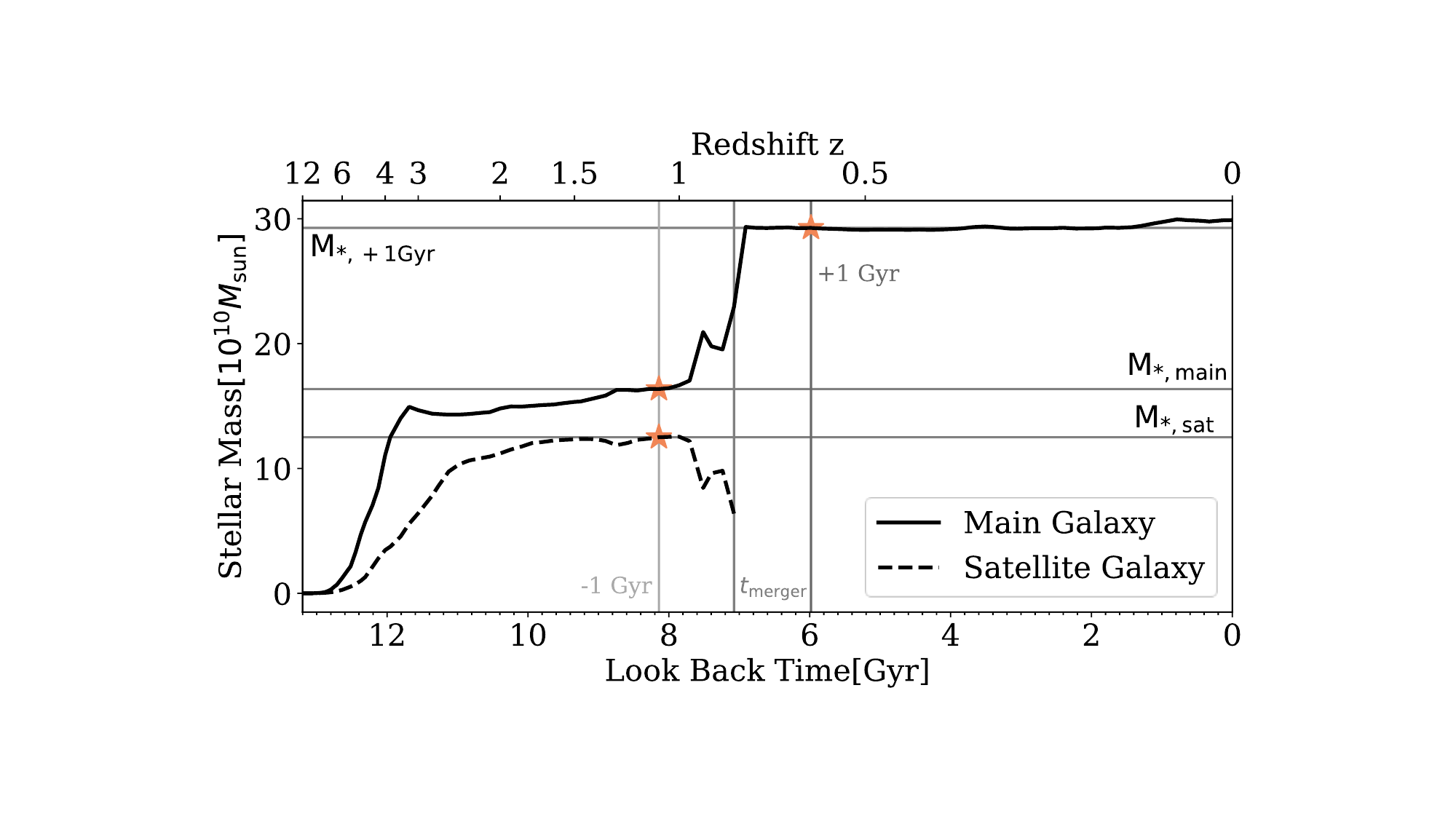}
\caption{{Stellar mass evolution of a merger pair. The solid black line and dashed black line represent the mass growth of the main galaxy and satellite galaxy, respectively.} The three horizontal lines mark the definition of main galaxy mass ($M_{*,\rm main}$), satellite galaxy mass ($M_{*,\rm sat}$), and remnant galaxy mass ($M_{*, \rm + 1Gyr}$).
The three gray vertical lines represent the merger time ($t_{\rm merger}$), 1 Gyr before ($t_{\rm merger} -1$ Gyr), and 1 Gyr after ($t_{\rm merger} +1$ Gyr) the merger. 
}
\label{fig:merger_tree}
\end{figure}

\begin{figure*}
\includegraphics[width=18cm]{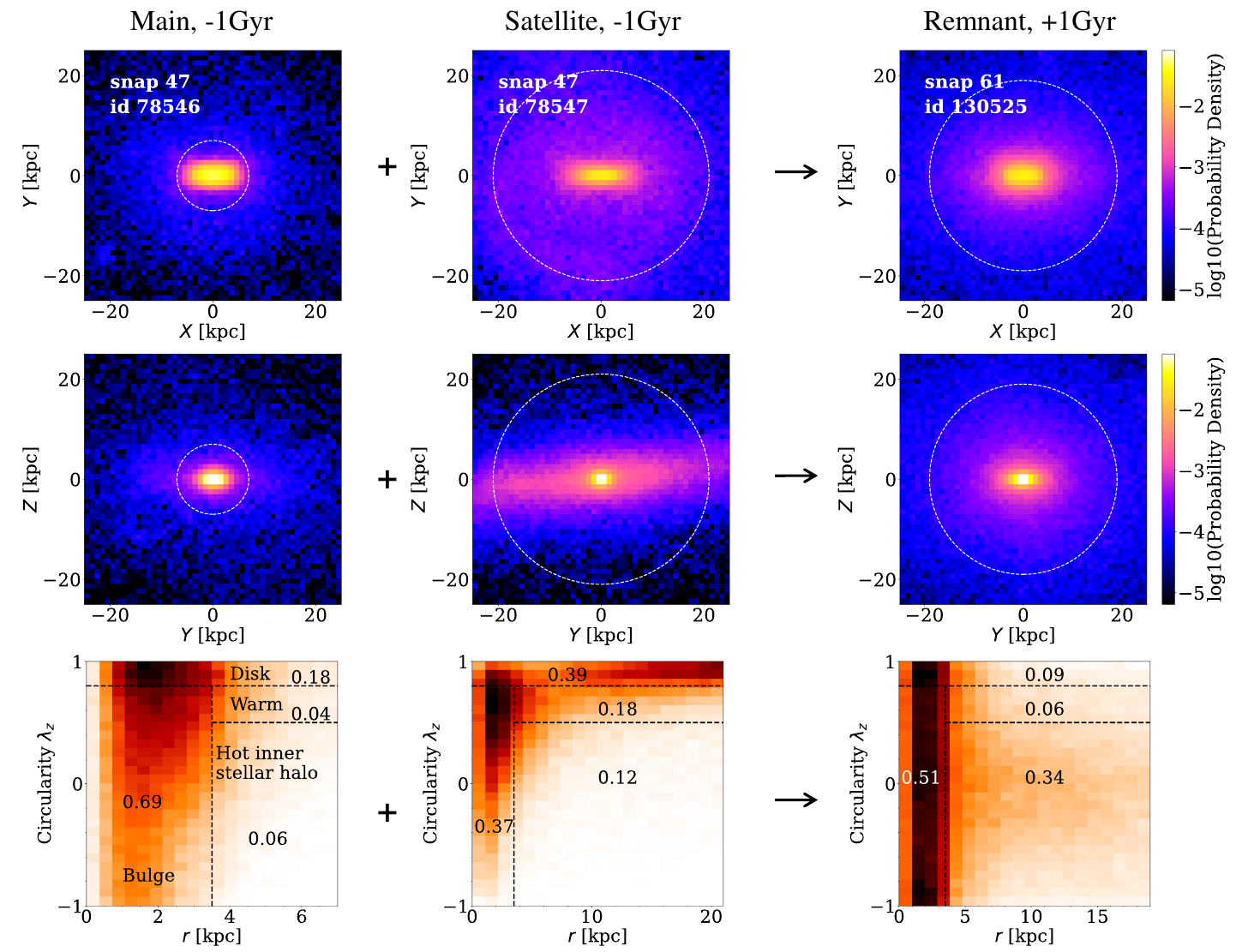}
\caption{Decomposition of galaxy structure adopted in this paper. We show the main progenitor, the satellite, and the remnant galaxy of a merger pair as the same one shown in Fig. \ref{fig:merger_tree}. The top  and the middle row show galaxies image projected on the X-Y and Y-Z plane, corresponding to face-on view and edge-on view, respectively. Columns from left to right are the main galaxy, satellite galaxy and remnant galaxy, respectively, with snapshot and ID shown in each panel of the top row. The part enclosed by the white dotted circle indicates two times of half-mass radius. The bottom row displays corresponding stellar orbit distribution within the white dotted circle as well as decomposition method based on the phase space of $\lambda_z$ versus r. The components of disk, warm, bulge, hot inner stellar halo, and the corresponding fractions are labeled in the panel. The merger destroys the disks of progenitor galaxies and causes an increase in the fraction of stars in the hot inner stellar halo of the remnant.
}
\label{fig:decomposition}
\end{figure*}

Depending on the volume and resolution, TNG contains three sub-projects: TNG50, TNG100, and TNG300, with the volume of the simulation box increasing and the resolution decreasing. The side lengths of the simulation boxes are 51.7, 110.7, and 302.6 Mpc, respectively. The mass resolutions of stellar and gas particles and cells are $8.5 \times 10^4$, $1.4 \times 10^6$, $1.1 \times 10^7$ \Msun, while the mass resolutions of dark matter particles are $4.5 \times 10^5$, $7.5 \times 10^6$, and $5.9 \times 10^7$ \Msun, for TNG50, TNG100, and TNG300, respectively. In this paper, we adopt TNG100, as its resolution (stellar and gas particles, and cells: $1.4 \times 10^6$ \,\Msun\,) remains sufficient to resolve different galaxy structures, such as disk, bulge, and inner stellar halo, for galaxies in our sample ($M_*>10^{10}$ \,\Msun\ at $z=0$). This is supported by the consistency of luminosity fractions in different dynamically decomposed components with TNG50 and the observations from CALIFA \citep{zhang2025diverse}. Moreover, there are still a sufficient number of galaxies in TNG100 to allow the selection of a representative sample.

In TNG simulations, halos and subhalos, and thus the basic properties of galaxies, are identified using the friends-of-friends (FoF) and \textsc{Subfind} algorithms \citep{springel2001populating, dolag2009mhd}. A subhalo with stellar particles is taken as a galaxy.
The evolutionary histories of all galaxies in the simulations can be followed via the merger trees, as constructed by the \textsc{Sublink-gal} code based on the baryonic component of subhalos \citep{rodriguez2015merger}. In this algorithm, each galaxy is assigned a unique descendant. We can identify, for each galaxy, its mergers with secondary galaxies with a well-defined ``infall'' time and that do not exist after ``coalescence'' as individually-identified \textsc{Subfind} objects. For each galaxy at a given time, we can also identify its satellites, namely, galaxies with a well-defined ``infall'' time that orbit or fly by around the galaxy before possibly merging with it.

In this work, we focus on the merger process of two galaxies, searching for such merger pairs through the merger tree.
In Fig. \ref{fig:merger_tree}, we present an example of a merger pair identified from the merger tree of subhalo 241153. We define the merger time ($t_{\rm merger}$) as the last snapshot where the satellite galaxy is individually identified, which is at snapshot 54 for this pair, with the corresponding look-back time of 7.24 Gyr. 

The mass and structure of galaxies are continuously changing throughout the merger process. In order to evaluate the change of structures by the merger, we have to properly choose the progenitor and remnant galaxies before and after the merger.
We uniformly take the progenitor galaxies at the snapshot 1 Gyr before the merger $t_{\rm -1 Gyr}$, when, for most cases, the internal structures of the progenitor galaxies have not been significantly influenced by each other. We thus define the stellar mass of progenitor galaxies at this snapshot as the satellite galaxy mass ($M_{*,\rm sat}$) and the main galaxy mass ($M_{*, \rm main}$), respectively.
For the remnant galaxy, we used the snapshot 1 Gyr after the merger($t_{\rm +1 Gyr}$), which is the time when the galaxy has typically settled down and reached dynamical equilibrium, and the remnant galaxy mass ($M_{*, \rm +1Gyr}$) has been defined accordingly.
By this definition, the remnant galaxy mass is close to or larger than the sum of the main galaxy and the satellite galaxy mass. 
We illustrate the definition of merger time ($t_{\rm merger}$), 1 Gyr before the merger ($t_{\rm -1 Gyr}$) and 1 Gyr after the merger ($t_{\rm +1 Gyr}$), stellar mass of the main galaxy ($M_{\rm *, main}$), satellite galaxy ($M_{\rm *, sat}$), and the remnant galaxy at 1 Gyr after the merger ($M_{\rm *,+1Gyr}$) in Fig. \ref{fig:merger_tree}. In the following sections, we  consider the merger orbit from $t_{\rm -1 Gyr}$ to $t_{\rm merger}$; this choice is aligned with previous studies \citep{zeng2021formation} as well.

\subsection{Orbital structure decomposition}
\label{sec:decomposition}

The internal stellar structure of a galaxy can be comprehensively described by its stellar orbit distribution. We can characterize the stellar orbit with two parameters, the radius, $r$, and the circularity, $\lambda_z$, where $\lambda_z$ is calculated as $\lambda_z=J_z/J_{\rm max}(E),$ with $J_z$ representing the z-component of the angular momentum of the orbit and $J_{\rm max}(E)$ representing the maximum angular momentum of a circular orbit with the same binding energy, E. The $\lambda_z$ ranges from -1 to 1.
A value of $\lvert \lambda_z \rvert \sim 1$ represents near circular orbits, $\lvert \lambda_z \rvert \sim 0$ indicates radial-motion dominated or box orbits, and $\lambda_z <0 $ corresponds to counter-rotation (CR) orbits. In principle, both $r$ and $\lambda_z$ are average values of many particles sampled along an orbit. For simulated galaxies, we have instantaneous 6D phase-space information of the particles at each snapshot, but lack the integration along orbits. 
In this paper, we use the instantaneous values of $r$ and $\lambda_z$ of the stellar particles.

We then decomposed a galaxy into different components based on the stellar orbit distribution. In this paper, we decompose a galaxy structure into four components: disk, warm, bulge, and hot inner stellar halo following \cite{zhu2022mass}. Specifically, the disk component refers to the region, where $\lambda_z$ > 0.8. The warm component encompasses the region where 0.8 > $\lambda_z$ > 0.5 and $r_{\rm cut}$ < r < $r_{\rm max}$. The bulge component corresponds to the region, where $\lambda_z$ < 0.8 and r < $r_{\rm cut}$. The hot inner stellar halo component includes the region where $\lambda_z$ < 0.5 and $r_{\rm cut}$ < r < $r_{\rm max}$. Here, $r_{\rm cut}$ = 3.5 kpc and $r_{\rm max}$ = 2$R_e$ for $R_e$ > 3.5 kpc, while $r_{\rm max}$ = 7 kpc otherwise. In this context, $R_e$ represents the half-mass radius of the galaxy, which is directly obtained from the TNG. 

In Fig. \ref{fig:decomposition}, we present the galaxy image and stellar orbit distribution for galaxies in a merger pair, including the progenitor galaxies at 1 Gyr before the merger and the remnant galaxy at 1 Gyr after the merger. We have rotated the galaxy so that the X-Y plane represents the face-on view and Y-Z plane corresponds to the edge-on view, with the orientation based on the principle eigenvector of the stellar moment of inertia tensor. From the stellar orbit distribution, we obtained  disk fractions of 0.18 for the main progenitor galaxy and 0.39 for the satellite galaxy. After the merger, the disk fraction decreases to only 0.09, where the disk fraction is defined as the mass of the disk($\lambda_z$>0.8) divided by the total stellar mass, without any radial restriction.
In the remnant galaxy, no obvious disk is presented, but a lot of stars reside in the hot inner stellar halo, accounting for 0.34 of the total stellar mass. The reshaping of galaxy structures by a merger can thus be quantitatively described by the change of these different structures.

We note that random motions of stars in the bulge may result in some of these stars also having high values of $\lambda_z$ \citep{aumer2013towards,marinacci2014formation}, which may affect our definition of the disk fraction, $f_{\rm disk}$. Thus, we   checked all our results using the more rigorous definition, where the disk fraction is computed as the fractional mass of stars with $\lambda_z > 0.8$ minus the mass of stars with $\lambda_z < -0.8$. This definition produces a systematic decrease in $f_{\rm disk}$, especially for galaxies with very low disk fractions. We note that none of our scientific conclusions are affected by this refinement.

\begin{figure}[h!]
\centering\includegraphics[width=9cm]{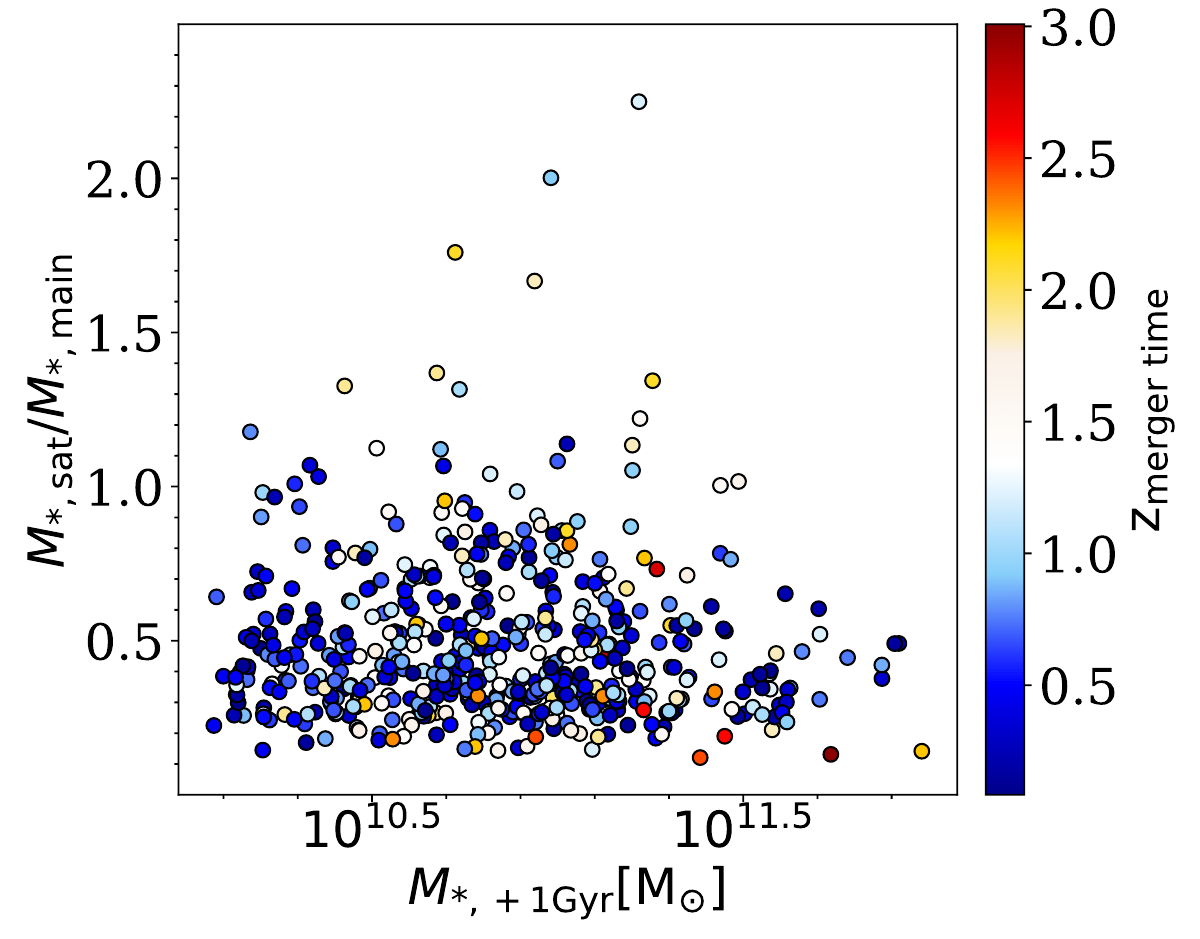}
\caption{Merger mass ratio ($M_{*,\rm sat}/M_{*,\rm main}$) vs. remnant galaxy mass ($M_{*,\rm +1Gyr}$) of merger pairs used in this paper. Each dot represents one merger pair and we have 531 merger pairs with a wide range of merger ratios. The color of the dots represent the redshift of the merger time.
} 
\label{fig:merger_sample}
\end{figure}

\subsection{Sample selection}
\label{s:selection}

We first searched for merger pairs in the merger trees of all central galaxies with $M_*>10^{10}$\,\Msun\, at $z=0$ in TNG100. Then we selected the merger pairs that satisfy the following criterion.
First, we chose the merger pairs with a mass ratio of more than 1/5, which allowed us to investigate the influence of different mass ratios on the reshaping of the galaxy morphology.
Here, the mass ratio refers to the ratio of stellar mass of the satellite galaxy and the main galaxy at $t_{\rm -1Gyr}$. Our definition differs from several other TNG simulation studies (e.g., \cite{rodriguez2016stellar}) that have sought to determine the merger ratio when the satellite galaxy attains its peak stellar mass. When applying the alternative definition to our analysis, we observed only minor variations in the results compared to our original approach, with no significant impact on our scientific conclusions.
We note that we define the main galaxy as the one identified as the first progenitor in \textsc{Sublink}, namely, the one with the "most massive history" behind it \citep{de2007hierarchical}. The satellite galaxy is defined as the one identified as the next progenitor. As a result, there are cases where the mass ratio exceeds 1.
Then we restrict the disk fraction of the main progenitor to be larger than 0.1, to exclude galaxies dominated by boxy orbits thus maintain the diversity of our sample. The disk fractions of the main progenitors in our final sample range from 0.1 to 0.65. Furthermore, we also examined the correlation between the two mass ratios,  finding that they exhibit a clear correlation centered along the 1:1 relation, with 66\% of the merger ratio displaying a difference of less than 0.1 and 83\% a difference of less than 0.2.

We visually inspected all the merger trees and exclude those incorrectly identified by \textsc{Sublink}.
We further checked the Y-Z projection (edge-on image) as well as the $\lambda_z$ - r phase space of all the progenitor galaxies at $t_{\rm -1 Gyr}$ and the remnant galaxies at $t_{\rm +1 Gyr}$. We manually excluded cases where the progenitor galaxies' structures have already been influenced at $t_{\rm -1 Gyr}$ or where the remnant galaxy has not yet reached dynamical equilibrium at $t_{\rm +1 Gyr}$ (i.e., exhibiting clear substructures upon visual inspection). This ensures a fair comparison of the "settled" galaxy structures before and after mergers. We show four of our manually excluded samples in Fig. \ref{fig:manually_excluded}.

We ultimately select a sample of 531 galaxy merger pairs, as shown in Fig. \ref{fig:merger_sample}. We have 162, 170, and 199 merger pairs in the mass ratio ranges of 1/5-1/3, 1/3-1/2, and 1/2-2, respectively. 

\begin{figure}
\includegraphics[width=8cm]{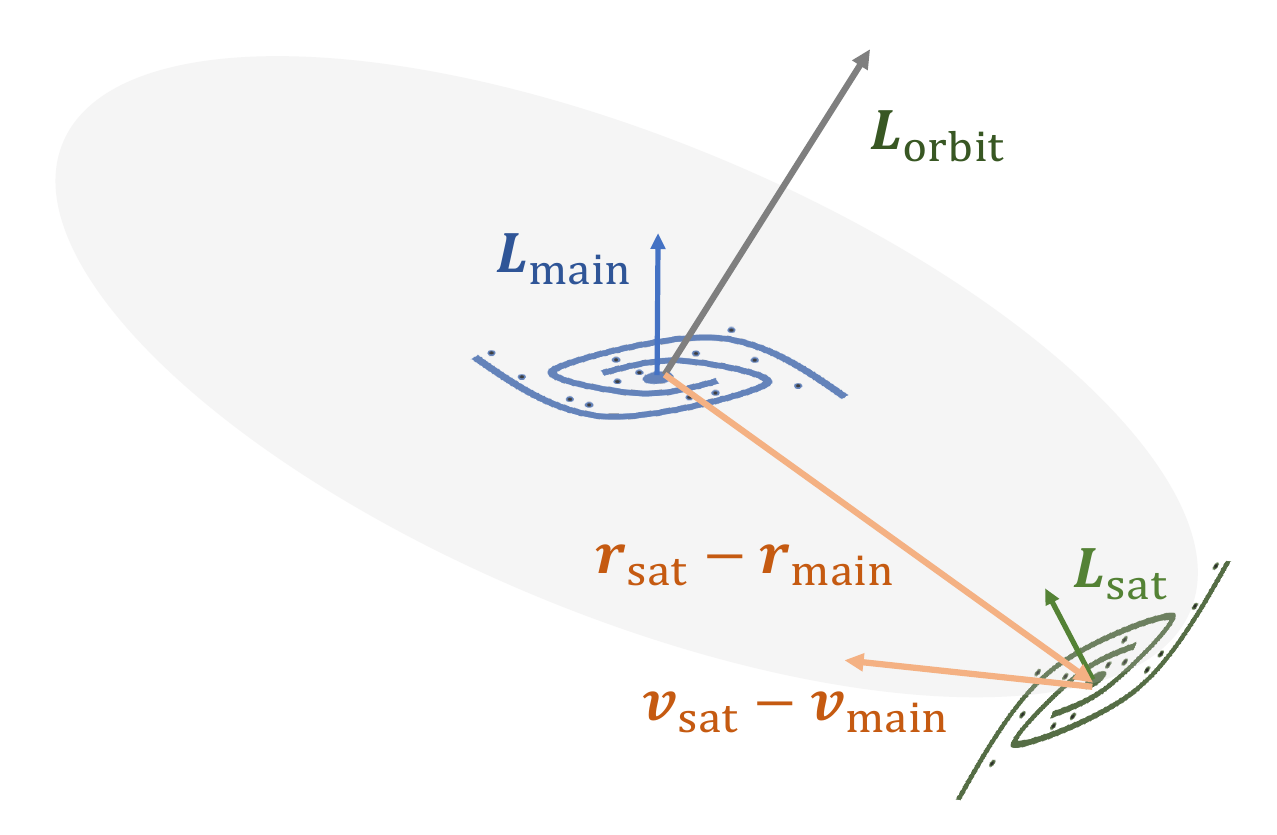}
\caption{A diagram illustrating the relative position of two galaxies during the merger. The blue plane represents the main galaxy, while the green plane represents the satellite galaxy and the gray plane represents the orbital plane, with corresponding colored vectors indicating the spin directions. The two orange vectors represent the position and velocity of the satellite galaxy relative to the main galaxy. This diagram is adapted from Figure 5 in \cite{martin2018role}.}
\label{fig:angle_define}
\end{figure}

\subsection{Definition of merger angles}

For a merger pair, we first computed the spin of the merging orbital plane by the cross product of the position- and velocity-vector of the satellite galaxy relative to the main galaxy, denoted as: 
\begin{equation}
\textbf{L}_{\rm orbit}=M_{sat}[(\textbf{r}_{\rm sat}-\textbf{r}_{\rm main})\times(\textbf{v}_{\rm sat}-\textbf{v}_{\rm main})]. 
\end{equation}
We defined the angle between the position- and velocity-vector (\anglerv) to characterize the merger orbit:
\begin{equation}
\cos(\theta_{\textbf{r},\textbf{v}}) = (\textbf{r}_{\rm sat}-\textbf{r}_{\rm main})\cdot(\textbf{v}_{\rm sat}-\textbf{v}_{\rm main})/(|\textbf{r}_{\rm sat}-\textbf{r}_{\rm main}| |\textbf{v}_{\rm sat}-\textbf{v}_{\rm main}|).
\end{equation}

The two galaxies could move close and apart during the merger. Thus, we would take \anglerv = $180^{\circ}$-\anglerv\, if \anglerv > $90^{\circ}$. By this definition, \anglerv = $0^{\circ}$ means that the satellite galaxy collides directly with the main galaxy in a radial orbit and \anglerv = $90^{\circ}$ means that the satellite galaxy spirals inward in a circular orbit. 

We then used the spin direction calculated by all the particles and cells in the main and satellite galaxies to represent their disk orientations, denoted as $L_{\rm main}$ and $L_{\rm sat}$. These spin directions are directly obtained from the spin parameter for each subhalo from the TNG catalogue. 
We defined the angle between the spin of the main galaxy and the orbital plane $\theta_{\textbf{L}_{\rm main} \textbf{L}_{\rm orbit}}$, with $\theta_{\textbf{L}_{\rm main},\textbf{L}_{\rm orbit}}=0$ indicating that the disk of the main galaxy is parallel to the orbital plane.  The angle between the spin of the satellite galaxy and the orbital plane was then $\theta_{\textbf{L}_{\rm sat},\textbf{L}_{\rm orbit}}$, with $\theta_{\textbf{L}_{\rm sat}, \textbf{L}_{\rm orbit}}=0$ indicating that the disk of the satellite galaxy is parallel to the orbital plane. The definition of these angles is summarized in Table \ref{tab:tab1}.

\begin{figure}[h!]
\centering\includegraphics[width=9cm]{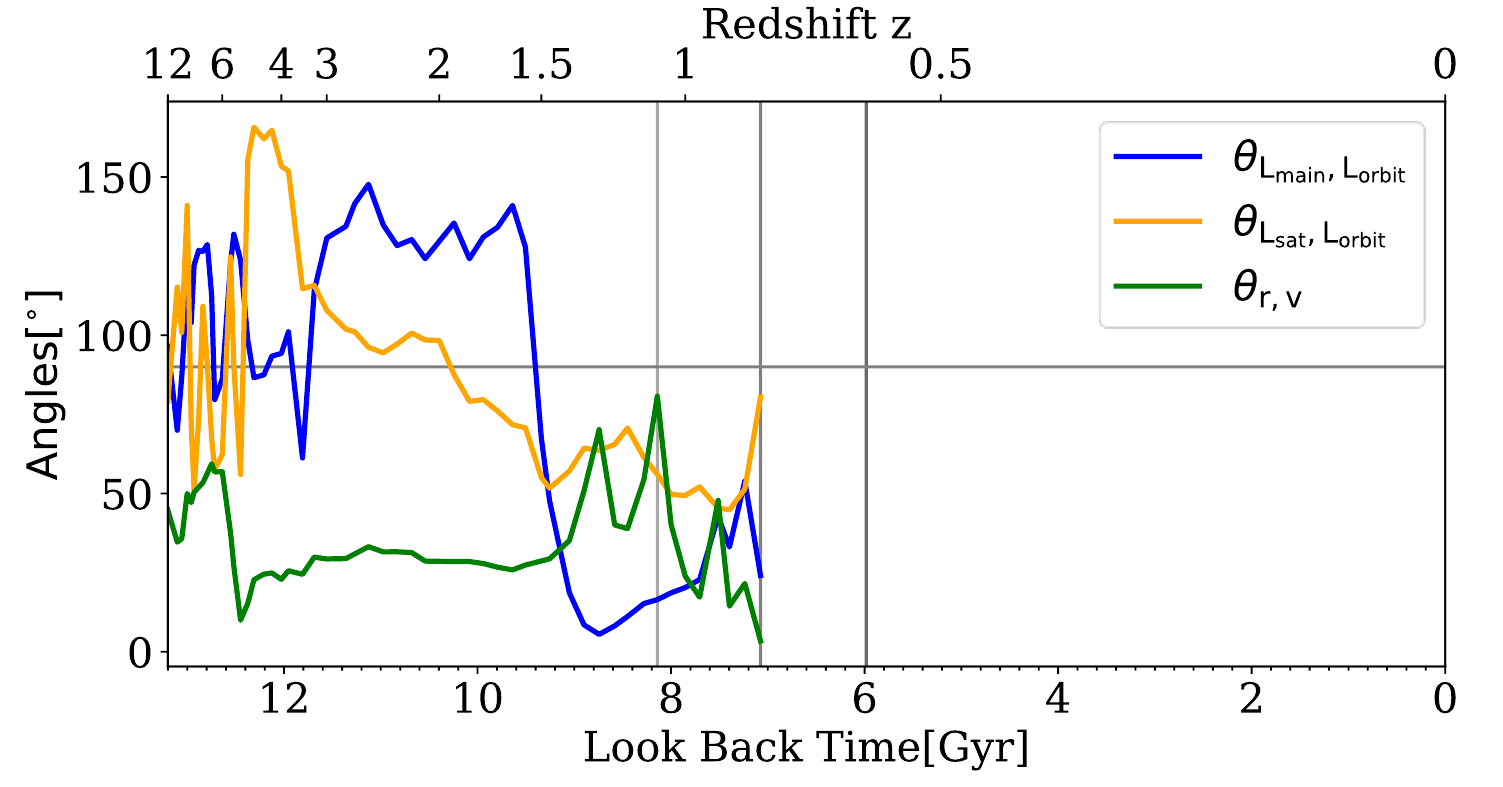}
\caption{Illustration of the evolution of the three angles before and during the merger. \anglemo\,, \,\angleso\,, and \,\anglerv\, are represented by blue, orange, and green lines, respectively. 
The three gray vertical lines represent the merger time ($t_{\rm merger}$) as well as 1 Gyr before ($t_{\rm merger} -1$ Gyr) and 1 Gyr after ($t_{\rm merger} +1$ Gyr) the merger, consistent with Fig. \ref{fig:merger_tree}. The gray horizontal line represents $90^{\circ}$. 
The illustrated merger is the same pair as shown in Figs.\ref{fig:merger_tree} and \ref{fig:decomposition}.} 
\label{fig:angle_tree}
\end{figure}

These angles undergo unstable variations during the merger, as shown in Fig. \ref{fig:angle_tree}. To better characterize the merger orbits, we calculated the average value of each angle across all snapshots within the 1 Gyr period leading up to the merger event and we took the average value in all the following analyses, denoted by $\langle \theta_{\textbf{L}_{\rm main},\textbf{L}_{\rm orbit}} \rangle$, $\langle \theta_{\textbf{L}_{\rm sat},\textbf{L}_{\rm orbit}} \rangle$ and \aanglerv.
After taking the average within the 1 Gyr period before the merger, we further folded $\langle \theta_{\textbf{L}_{\rm main},\textbf{L}_{\rm orbit}} \rangle$ and $\langle \theta_{\textbf{L}_{\rm sat},\textbf{L}_{\rm orbit}} \rangle$ to limit them within 0-90 degrees as follows:
\begin{equation}
\langle \theta_{\textbf{L}_{i},\textbf{L}_{\rm orbit}}' \rangle=
\begin{cases}
\langle \theta_{\textbf{L}_{i},\textbf{L}_{\rm orbit}\rangle},&{if \langle \theta_{\textbf{L}_{i},\textbf{L}_{\rm orbit}}\rangle <90^{\circ},}\\
180^{\circ}-\langle \theta_{\textbf{L}_{i},\textbf{L}_{\rm orbit} \rangle},&{if \langle \theta_{\textbf{L}_{i},\textbf{L}_{\rm orbit}}\rangle >90^{\circ}}.
\end{cases}
\end{equation}

\begin{figure*}
\centering\includegraphics[width=14cm]{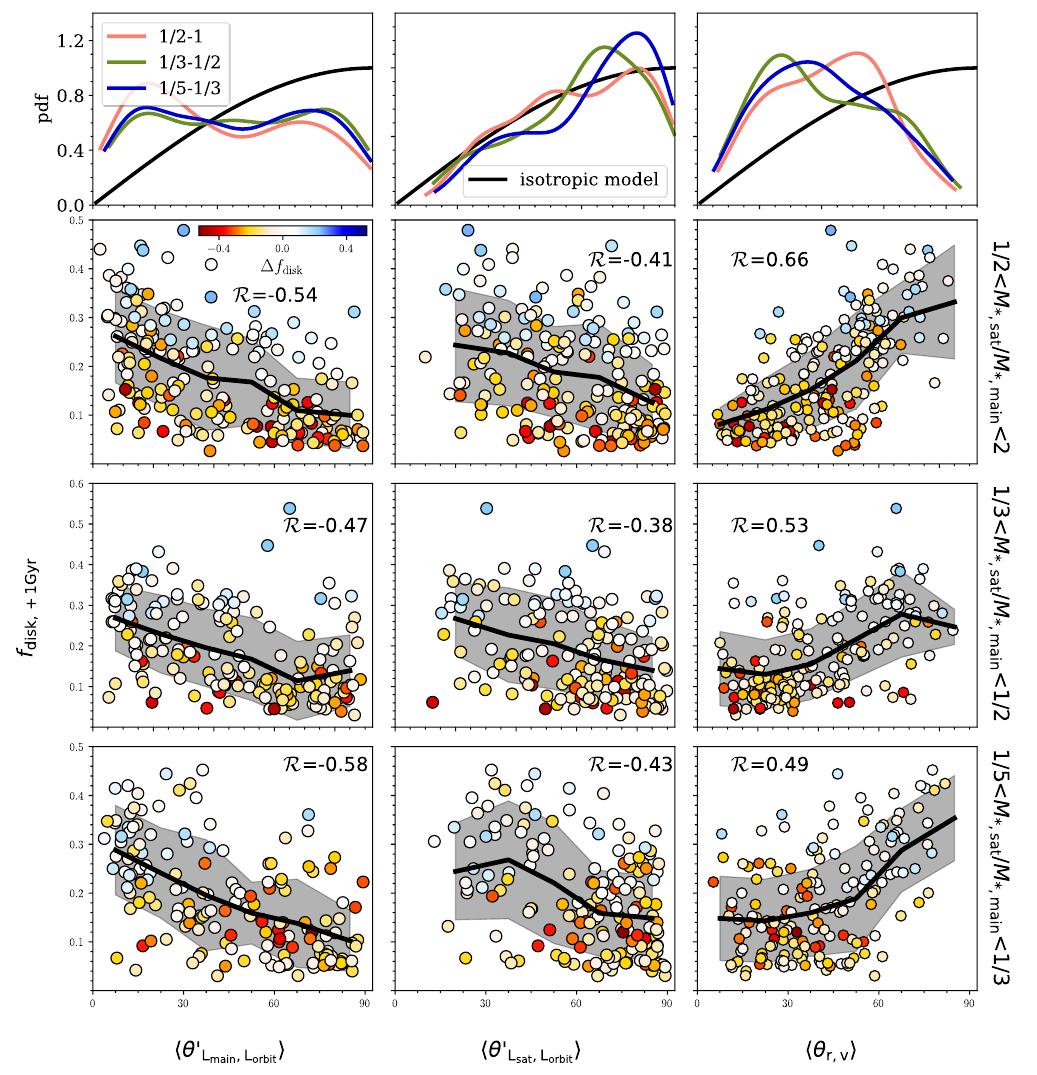}
\caption{Disk fraction in the remnant galaxy at 1 Gyr after the merger as a function of the three angles. The columns from left to right are $\langle \theta_{\textbf{L}_{\rm main},\textbf{L}_{\rm orbit}}' \rangle$, $\langle \theta_{\textbf{L}_{\rm sat},\textbf{L}_{\rm orbit}}' \rangle$, and \aanglerv. The rows from the second top to the bottom corresponds to mergers with mass ratio of 1/2 - 2, 1/3 - 1/2, and 1/5 - 1/3, respectively. Each dot represents one merger pair colored by $\Delta f_{\rm disk}$, with red dots denoting a decrease in the disk fraction and blue dots denoting an increase in the disk fraction compared to the main progenitor galaxy at 1 Gyr before the merger. In each panel, the black solid line represents the running median of $f_{\rm disk,+1Gyr}$ and the gray shadow regions indicate the $\pm 1 \sigma$ scatter. The correlation coefficients between $f_{\rm disk,+1Gyr}$ and angles are shown in each panel. The top row illustrates the probability density function (PDF) of the angles. The red, green, and blue lines represent mass ratio between 1/2 and 2, 1/3 and 1/2, 1/5, and 1/3, respectively, all smoothed using Gaussian kernel density estimation (KDE). The black line represents the PDF expected from an isotropic model. 
} 
\label{fig:fdiskA}
\end{figure*}

\begin{table}[!ht]
\begin{threeparttable}
\centering
\caption{Definition of orbital and spin angles.}
\label{tab:tab1}
\begin{tabular}{l p{1.8cm} p{1.8cm}}
\toprule
\toprule
Angle &  Vector1 & Vector2 \\ \midrule
\anglemo & $\textbf{L}_{\rm main}$ & $\textbf{L}_{\rm orbit}$ \\
\angleso & $\textbf{L}_{\rm sat}$ & $\textbf{L}_{\rm orbit}$ \\
\anglerv & $\textbf{r}_{\rm sat}-\textbf{r}_{\rm main}$ & $\textbf{v}_{\rm sat}-\textbf{v}_{\rm main}$ \\
\bottomrule
\end{tabular}
\end{threeparttable}
\end{table}

\begin{figure*}
\centering\includegraphics[width=18cm]{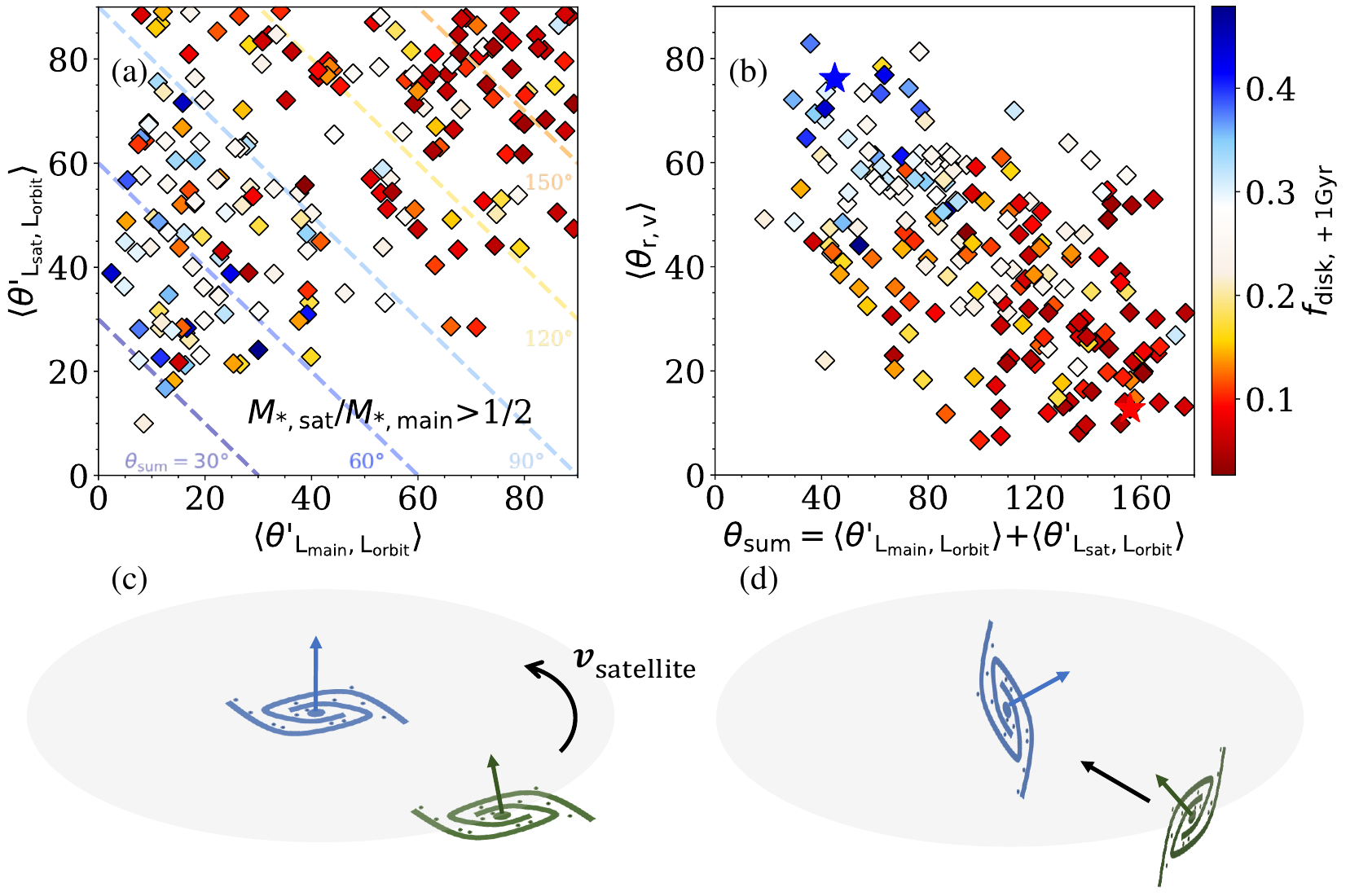}
\caption{Disk fraction in the remnant galaxy ($f_{\rm disk, +1Gyr}$) as a function of $\langle \theta_{\textbf{L}_{\rm main},\textbf{L}_{\rm orbit}}' \rangle$, $\langle \theta_{\textbf{L}_{\rm sat},\textbf{L}_{\rm orbit}}' \rangle$, and \aanglerv\, for the mergers with mass ratio over 1/2. 
The dots in panels (a) and (b) are colored with $f_{\rm disk, +1Gyr}$. The dashed lines in panel (a) denote $\theta_{\rm sum}=30^{\circ}, 60^{\circ}, 90^{\circ}, 120^{\circ}$, and $150^{\circ}$ from the bottom-left to the top-right. Panels (c) and (d) illustrate two extreme cases result in high and low disk fractions in the remnant galaxy. In panel (c), both the main progenitor galaxy and the satellite galaxy are aligned parallel to the merger orbital plane, and their orbital configuration is spiral-in, leading to a high disk fraction in the remnant galaxy.
In panel (d), both the main progenitor galaxy and the satellite galaxy are aligned perpendicular to the merger orbital plane and with a head-on orbit, resulting in a very low disk fraction in the remnant galaxy.
For the two typical scenarios represented by panels (c) and panel (d), two samples are selected for display in Appendix~\ref{subsec:typical sample}. These two samples are indicated in panel (b) by a blue star in the upper left corner and a red star in the lower right corner.
}
\label{fig:fdiskAA}
\end{figure*}

\section{Results}
\label{sec:results}

\subsection{Distribution of merger angles}
\label{sss:angleA}

We divided the 531 merger pairs into three groups with a merger mass ratio of 1/2-2, 1/3-1/2, and 1/5-1/3. In the upper row of Fig. \ref{fig:fdiskA}, we present the probability distribution function (PDF) of the three angles, along with the theoretical PDF expected from an isotropic model, as expressed by Equation 5 in \cite{cai2013distributions}:
\begin{equation}
p(\theta) = \rm sin(\theta),\quad 0<\theta<90^{\circ} \quad \text{for 3D cases.}
\end{equation}

We find that the distribution of \,\afangleso\ is consistent with an isotropic model, indicating that the merger orbit has little influence on the orientation of satellite galaxies and vice versa.
\,\afanglemo\ is almost uniformly distributed between 0° and 90°, for all mergers with different mass ratios. Compared to the expectation from an isotropic model, there are more mergers where the merger orbital plane aligns with the disk plane of the main progenitor galaxy. This could result from the anisotropic accretion of the filaments \citep{welker2014mergers} or from the tilt of the satellite orbit due to the tidal torque of the disk plane, especially for minor mergers \citep{read2009dark}.

The distribution of \,\aanglerv\ shows
a strong preference for head-on collisions on radial orbits over spiral-in mergers on tangential orbits. This could be explained by \cite{vasiliev2022radialization}, which illustrates that the orbits of satellite galaxies have a tendency to radialize during merger, especially for mergers with a high satellite mass (mass ratio>1/10). 

Furthermore, our analysis reveals that in all three PDFs, distributions are largely similar irrespective of merger mass ratio. As demonstrated by \cite{li2020orbital}, the infall angle \,\anglerv\ between satellite and host halos is completely determined by the satellite halo’s velocity, the host mass, $\upsilon$, and sub-to-host halo $\xi$. \cite{vasiliev2022radialization} further established that orbital radialization depends on the host galaxy's cusp slope $\gamma$, initial circularity $\eta$, and satellite-to-host mass ratio. Given that we do not have any environmental selections in our sample, it is reasonable to assume that \,\aanglerv\ shows similar distribution across the relatively narrow mass ratio range of 1/5-2. The correlation between orbital configurations and galaxy orientations  (as shown in Fig. \ref{fig:fdiskAA}) further supports extending this interpretation to \,\afanglemo\ and \afangleso .

\begin{figure*}
\centering\includegraphics[width=18cm]{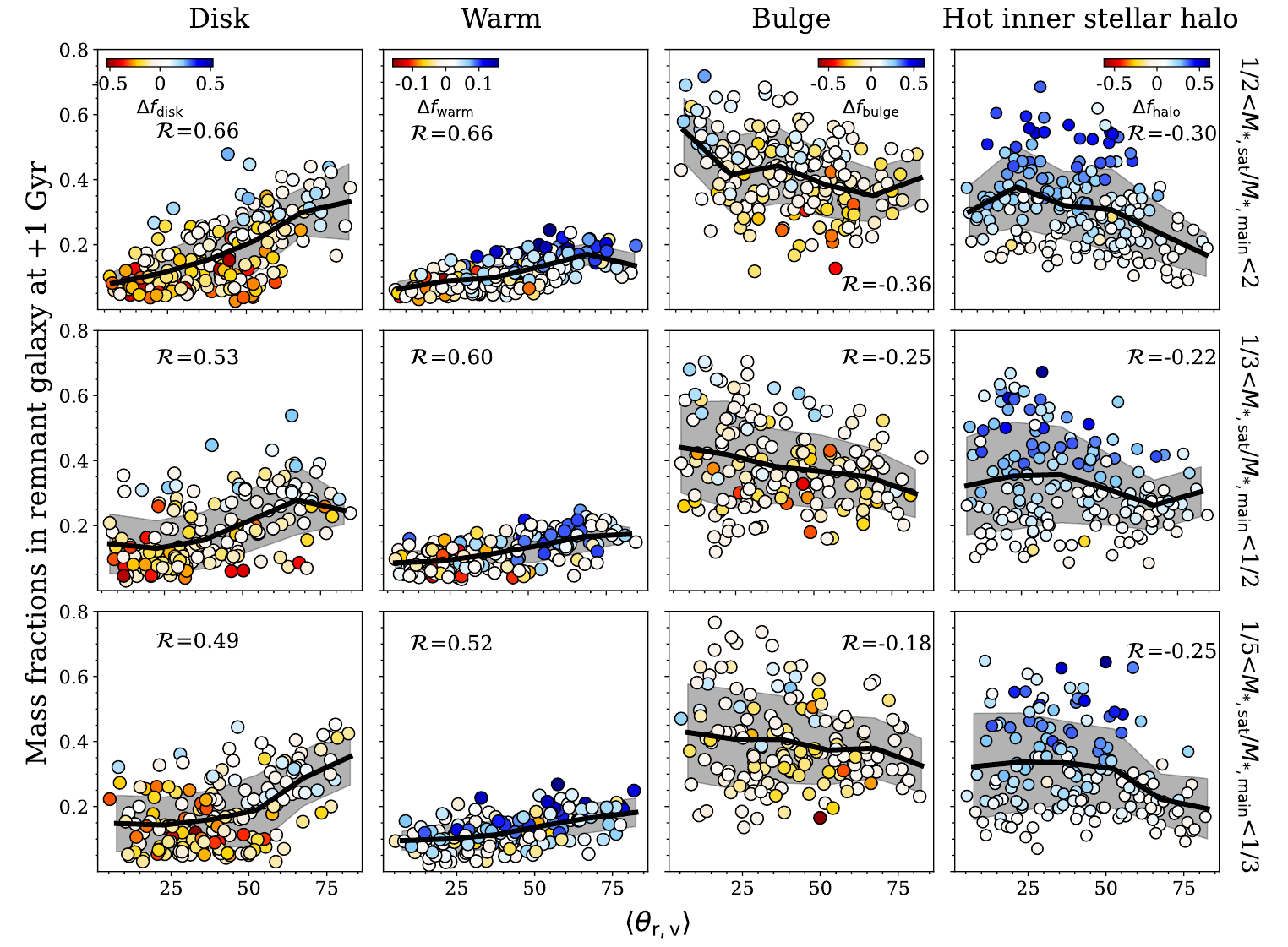}
\caption{Mass fraction of different orbital components in the remnant galaxy at 1 Gyr after the merger as a function of \aanglerv. Columns from left to right: Disk, warm component, bulge, and hot inner stellar halo. Rows from top to bottom: Mergers with mass ratios of 1/2-2, 1/3-1/2, and 1/5-1/3, respectively. In each panel, the colors indicate the change in the mass fraction of each component compared to that in the main progenitor galaxy at 1 Gyr before the merger, as illustrated by the colorbar. The solid black line represents the running median and the gray shadow region denotes the $1\pm \sigma$ scatter. The correlation coefficient $\mathcal{R}$ is calculated and labeled in each panel.  
}
\label{fig:angle_4comps}
\end{figure*}

\begin{figure*}
\centering\includegraphics[width=18cm]{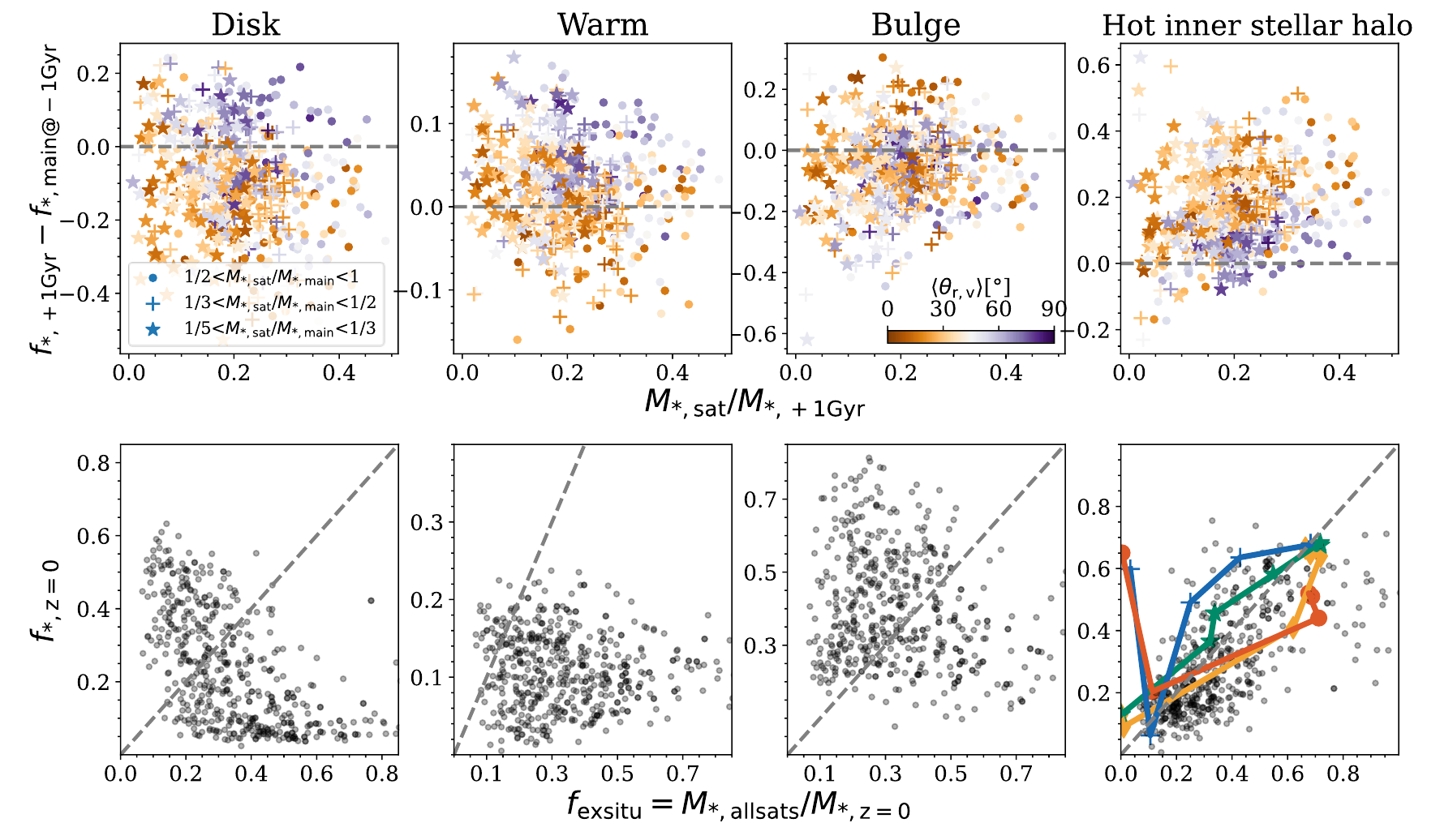}
\caption{Top:  Change in the mass fractions of disk, warm, bulge, and hot inner stellar halo vs. $M_{*,\rm sat}/M_{*,\rm +1 Gyr}$ for all the merger pairs.
The colors of the top panels represent \aanglerv. The  circle, plus, and star symbols denote mass ratios of $1/2-2$, $1/3-1/2$, and $1/5-1/3$, respectively. The gray dashed lines indicate no change in mass fractions.
Bottom: Mass fraction of different components in galaxies at $z=0$ vs. the total ex situ stellar mass fraction of the galaxy.
The gray dashed lines represent a 1:1 relationship. Since each galaxy may have experienced multiple mergers, we illustrate the change in $f_{\rm halo}$ after each merger with four galaxies, colored by blue, green, orange, and red. The adding-up of multiple mergers results in a good correlation between $f_{\rm halo, z=0}$ and the total ex situ stellar mass.
}
\label{fig:fdiskAAA}
\end{figure*}

\subsection{The dependence of disk survival on merger angles}
\label{sss:fdiskA}

For each merger pair, we computed the disk fraction in the main progenitor galaxy at 1 Gyr before merger ($f_{\rm disk, -1Gyr}$) and in the remnant galaxy at 1 Gyr after merger ($f_{\rm disk, +1Gyr}$). Then, we calculated the change in disk fraction with $\Delta f_{\rm disk} = f_{\rm disk, +1Gyr} - f_{\rm disk, -1Gyr}$.

In Fig. \ref{fig:fdiskA}, we show the correlations between the disk fraction in the remnant galaxy $f_{\rm disk, +1Gyr}$ and the three angles that describe the merger orbital configuration. In comparison, the correlation between $f_{\rm disk,+1Gyr}$ and the unfolded angles $\langle \theta_{\textbf{L}_{\rm main},\textbf{L}_{\rm orbit}} \rangle$ and $\langle \theta_{\textbf{L}_{\rm sat},\textbf{L}_{\rm orbit}} \rangle$ is shown in Appendix \ref{s:appendixb}. For major mergers with a mass ratio of 1/2-2, the disk fraction in the remnant galaxy $f_{\rm disk, +1Gyr}$ is negatively correlated with \afanglemo,\  with a Pearson correlation coefficient of $\mathcal{R}=-0.54$, and \afangleso , with $\mathcal{R}=-0.41$. This indicates that the $f_{\rm disk, +1Gyr}$ in the remnant is higher when either the disk plane of the main progenitor or the satellite galaxy aligns with the merger orbital plane (\afanglemo $\sim 0$ or \afangleso $\sim 0 $). This result is consistent with the findings of \cite{sotillo2022merger}, who showed that disks are more likely to survive in mergers with prograde orbits, where the spin of the satellite galaxy is aligned with the merger orbit.
We found a stronger correlation between $f_{\rm disk, +1Gyr}$ and the angle describing the merger orbit \,\aanglerv\, with $\mathcal{R} = 0.66$. This suggests that $f_{\rm disk, +1Gyr}$ tends to be higher when the satellite galaxies merge in tangential orbits (\aanglerv $\sim 90^{\circ}$ ), consistent with the results in \cite{zeng2021formation}.
The correlations mentioned above exhibit a relatively similar behavior across mergers with different mass ratios.

Most galaxies have disks at least partially destroyed during the merger, leading to a decreased disk fraction in the remnant galaxies. However, some galaxies show no change in the disk fraction or even an increase. Specifically, the fraction of galaxies with the disk fraction increased after merger is 26\%, 29\%, 25\% for merger ratios of 1/2-2, 1/3-1/2, and 1/5-1/3, respectively.

We have seen that the survival of disks shows a strong dependence on merger angles, especially for galaxies that have experienced major mergers. Here, we aim to further explore the types of galaxies that have the highest disk fractions by considering the full merger orbital configurations combining the three angles.

In panel (a) of Fig. \ref{fig:fdiskAA},  we show the dependence of $f_{\rm disk, +1Gyr}$ on \afanglemo\  versus \afangleso, focusing on major mergers.
The disk fraction in the remnant galaxies increases from the upper right to the lower left. The highest disk fractions result from merger cases where both the disk plane of the main progenitor galaxy and the satellite galaxy aligned with the plane of merger orbit (\afanglemo $\sim 0$ and \afangleso $\sim 0$), as illustrated in panel (c). The lowest disk fractions result from merger cases in which both the disk plane of the main progenitor galaxy and the satellite galaxy are almost perpendicular to the plane of merger orbit (\afanglemo\  $\sim 90^{\circ}$ and \afangleso\  $\sim 90^{\circ}$), as illustrated in panel (d).
Therefore, we introduced a new parameter, $\theta_{\rm sum}$, defined as $\theta_{\rm sum}$ = \afanglemo + \afangleso, which increases from the lower left to the upper right in panel (a) and exhibits a stronger correlation with $f_{\rm disk, +1 Gyr}$.

In panel (b) of Fig. \ref{fig:fdiskAA},  we further show that the orbital configuration (\aanglerv) and the orientations of the merging galaxies ($\theta_{\rm sum}$) are strongly correlated. The mergers resulting in the highest $f_{\rm disk, +1 Gyr}$ are those with the lowest $\theta_{\rm sum}$ and, at the same time, are merging on nearly circular orbits (\aanglerv $\sim 90^{\circ} $) (panel (c)); while the mergers resulting in the lowest $f_{\rm disk, +1 Gyr}$ are those with the highest $\theta_{\rm sum}$ and, at the same time, are merging on radial orbits (\aanglerv $\sim 0 $) (panel (d)). The evolution of merger orbits and galaxy structures of these two typical cases are further shown in Appendix~\ref{subsec:typical sample}. We further examined the results presented in Fig. \ref{fig:fdiskAA} for mergers with lower mass ratios, specifically those with mass ratios ranging from 1/3 to 1/2 and 1/5 to 1/3. We ultimately found that they exhibit a trend resembling that seen for major mergers.

\subsection{The effects on formation of different structures}
\label{sss:4comps}

Galaxies consist of several different structures. We have dynamically decomposed each galaxy into a cold disk, a warm component, a hot compact bulge, and a hot inner stellar halo (as described in Section \ref{sec:decomposition}). In the previous section, we describe  the three angles we defined, whereby the angle describing the merger orbit (\aanglerv) has the strongest effects on the survival of the disk. Here, we further investigate its effects on the formation of the different structures in the remnant galaxies. 

In Fig. \ref{fig:angle_4comps}, we show the correlations between the merger angle \,\aanglerv\ and the mass fraction of disk, warm component, bulge, and hot inner stellar halo in the remnant galaxy at 1 Gyr after the merger. The color  distinguishes the change of mass fractions compared to the main progenitor galaxy at 1 Gyr before the merger ($\Delta f_{*} = f_{*,\rm +1 Gyr} - f_{*,\rm main @ -1 Gyr}$). 

First of all, we find that for the merger with all mass ratios, the mass fraction of the disk component mainly decreases; the mass fraction of the warm component takes only a small fraction of the galaxies and largely increases; the bulge fraction increases in about half of the cases and decreases in the other half; and the hot inner stellar halo mass fraction nearly always increases.

There are relatively strong positive correlations between \,\aanglerv\ and the mass fractions of the disk and warm components in the remnant galaxies, with Pearson's correlation coefficients of $\mathcal{R}(\langle \theta_{\textbf{r},\textbf{v}} \rangle, f_{\rm disk,+1Gyr}) = 0.66$ and $\mathcal{R}(\langle \theta_{\textbf{r},\textbf{v}} \rangle, f_{\rm warm,+1Gyr}) = 0.66$ for the major mergers. The correlations between $\langle \theta_{\textbf{r},\textbf{v}} \rangle$ and the mass fractions of the bulge and hot inner stellar halo components are much weaker, with Pearson's correlation coefficients of $\mathcal{R}(\langle \theta_{\textbf{r},\textbf{v}} \rangle, f_{\rm bulge,+1Gyr}) = -0.36$, and $\mathcal{R}(\langle \theta_{\textbf{r},\textbf{v}} \rangle, f_{\rm halo,+1Gyr}) = -0.30$ for the major mergers. We note that all of the above correlations are weaker in minor mergers.

Due to the dependence of galaxy structure formation on the merger angles, especially \aanglerv, it is difficult to find a structure that can be used as a universal indicator of the galaxy merger history (i.e., the mass of the merged satellite). In the top panels of Fig. \ref{fig:fdiskAAA}, we show the change in mass fractions of different components ($\Delta f_{*}$) versus the mass contribution of satellite galaxies to the remnant galaxies for the 531 merger events. Overall, we find that the change of mass fractions of each component is not significantly related to the mass fraction of satellite galaxies; for individual mergers, the structure of the remnant is affected by multiple factors, including the merger mass ratio and merger orbital configurations, resulting in significant scatter. 

The 531 mergers are selected along the merger history of 496 descendant galaxies at $z=0$. Each of the massive galaxies at $z=0$ has experienced multiple mergers in history. 
In the bottom of Fig. \ref{fig:fdiskAAA}, we present the mass fraction of each component in the galaxy at $z=0$ versus the ex situ stellar mass fraction of the galaxy; here, the latter is defined as the ratio of the mass of all the accreted satellite galaxies from mergers with mass ratio $>1/10$ to the total stellar mass of the galaxy at $z=0$ ($f_{\rm exsitu} = M_{*,\rm allsats}/M_{*,\rm z=0}$). We find that there is no obvious correlation between $f_{\rm exsitu}$ and the mass fraction of the bulge and warm components, a weak negative correlation with the disk fraction and, in particular, a strong positive correlation between $f_{\rm exsitu}$ and the hot inner stellar halo component. These findings are consistent with \cite{zhu2022mass}.

The scatter between the mass fraction of different components and accreted satellite mass fraction is averaged by adding-up multiple mergers. As illustrated by four cases (the colored lines in the bottom right panel), although they deviate greatly from the 1:1 line initially, they gradually converge back toward the vicinity of the line due to the averaging effect. This convergence occurs because the fraction of the hot inner stellar halo ($f_{\rm halo}$) almost always increases after each merger.

\section{Discussion}
\label{sec:discussion}

\begin{figure*}
\centering\includegraphics[width=18cm]{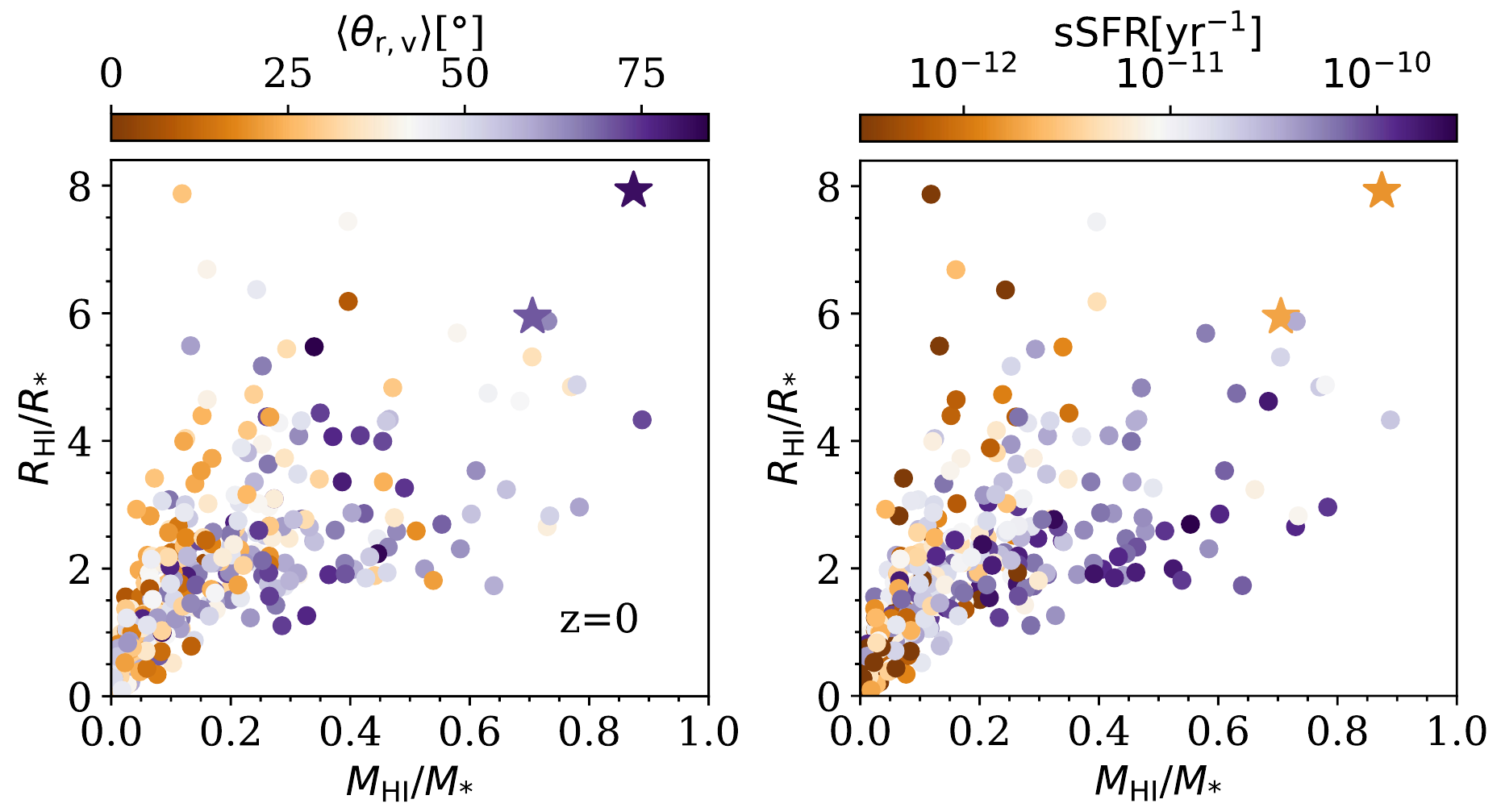}
\caption{$R_{\rm HI}/R_*$ vs. $M_{\rm HI}/M_*$ in the remnant galaxies at z=0 for our sample, colored by \aanglerv (left) and specific star formation rate (right).
In both panels, two galaxies marked by stars exhibit RR galaxies with high values of $M_{\rm HI}/M_*$ and $R_{\rm HI}/R_*$, yet their star formation appears to be quenched.
}
\label{fig:HI mass}
\end{figure*}

We investigate how the merger orbital configurations influence the stellar structure in the remnant galaxies. Overall, 59\% of these mergers are gas-rich with $M_{\rm HI}/M_{*}>0.1$. Here, we briefly check how the merger configurations affect the gas distribution in the remnant galaxies. 

A class of rare galaxies that are red, but \HI-rich (RR), was recently  reported by \cite{li2024existence}. These galaxies are red, indicating a fully-quenched star formation state, while they also have a high $M_{\rm HI}/M_*$.
Additionally, they show a relatively large \HI-to-optical radius ratio, which is approximately four times the average ratio observed in the normal red galaxy sample. 
The author suggested that this type of galaxy may inherit the high spin of their host dark-matter halos, while the external origin of their massive \HI gas content includes accretion and/or mergers.

From the remnants of the merger pairs analyzed in this study, we identified two simulated counterparts similar to the RR galaxies, as shown in Fig. \ref{fig:HI mass}.
These galaxies exhibit high values of $R_{\rm HI}/R_*$ and $M_{\rm HI}/M_*$, yet they are quenched (with sSFR<$10^{-11} yr^{-1}$). 
Both galaxies have experienced spiral-in mergers on highly tangential orbits, with \aanglerv = $81^{\circ}$ and $70^{\circ}$, respectively. We speculate that the spiral-in mergers could leave \HI gas behind along their merger trajectories, yielding enlarged \HI radius and \HI mass. However, the substantial residual angular momentum maintains this gas at the larger radius of the galaxy, leading to lower \HI density and consequent reduced sSFR. We suggest that merger on a highly tangential orbit offers an alternative explanation for the origin of the high angular momentum of \HI gas, thereby providing another formation pathway for  RR galaxies.

\section{Summary}
\label{sec:summary}

We selected 531 merger pairs with mass ratios greater than 1/5 and the remnant galaxies with stellar mass of $M_*>10^{10}$ \,\Msun\ at $z=0$ 
in TNG100. We adopted a uniform orbital structure decomposition method to describe the structures for galaxies before and after the merger, for all the 531 merger pairs. The galaxies are decomposed into four components: disk ($\lambda_z>0.8$), warm component ($0.8>\lambda_z>0.5$ and $r_{\rm cut}<r<r_{\rm max}$), bulge ($\lambda_z<0.8$ and $r<r_{\rm cut}$), and hot inner stellar halo ($\lambda_z<0.5$ and $r_{\rm cut}<r<r_{\rm max}$, where $r_{\rm cut}$ = 3.5 kpc, $r_{\rm max}$ = 2$R_e$ for $R_e$ > 3.5 kpc, and $r_{\rm max}$ = 7 kpc otherwise), each representing a unique dynamical structure. We comprehensively describe the merger orbital configurations defining three angles: \aanglerv\ describing the relative moving orbits of the merger pairs, \afanglemo\ describing the angle between the main galaxy spin and the orbital plane, and \afangleso\ describing the angle between the satellite galaxy spin and the orbital plane. We systematically studied the effects of merger orbits on galaxy structure formation. Our main results are as follows:

\begin{enumerate}

\item The merger orbital configurations of our sample deviate from isotropic distributions. In general, the spin of the main galaxy tends to align with that of the merger orbital plane: we have 58\% and 37\% mergers on radial orbits with \afanglemo<$45^{\circ}$ and \afanglemo<$30^{\circ}$, compared to 29\% and 13\% expected from an isotropic distribution, respectively. 
Additionally, the merging galaxies are more aligned with the merger orbital plane for spiral-in orbits than head-on collision orbits: the average value of \afanglemo+\,\afangleso\ is $116^{\circ}$ for mergers on radial orbits with \,\aanglerv\ $<45^{\circ}$, while it is $81^{\circ}$ for mergers on spiral-in orbits with \,\aanglerv\ $>45^{\circ}$ for major mergers (stellar mass ratio between 1/2 and 2).

\item The merger orbital configuration significantly affects the structure of the remnant galaxy. For mergers on spiral-in orbits, the disk planes of the two merging galaxies tend to be aligned with the orbital plane, leading to higher fractions of disk and warm components and lower fractions of
the bulge and hot inner stellar halo components in the remnant galaxy; for mergers on head-on collision orbits, the disk planes of the two galaxies tend to be perpendicular to the orbital plane, leading to lower fractions of disk and warm components and higher fractions of bulge and hot inner stellar halo in the remnant galaxy. For major mergers, the correlation coefficients of $f_{*,\rm +1Gyr}$ and \,\aanglerv\ are 0.66, 0.66, -0.36, and -0.30 for disk, warm component, bulge, and hot inner stellar halo, respectively.

\item Mergers can lead to either an increase or decrease in the disk and bulge mass fraction in
the remnant compared to the progenitor galaxy, depending on the merger orbital configurations, but mergers almost always cause an increase
in the hot inner stellar halo. The average effect of multiple mergers experienced by galaxies at z=0 results in a strong correlation between the fraction of the hot inner stellar halo and the total ex situ stellar mass fraction for galaxies at $z=0$. This is consistent with \cite{zhu2022mass}. 

\item Spiral-in mergers could be one of the origins for the rare red but \HI-rich (RR) galaxies. Two RR galaxies are found in the remnants of our merger pairs at $z=0$, both having undergone mergers with spiral-in orbits (\aanglerv = $81^{\circ}$ and $70^{\circ}$).

\end{enumerate}

We show that the merger orbital configuration significantly affects the structural evolution of galaxies. Therefore, we have to be cautious when using galaxy structures as an indication of galaxy merger histories, as neither the disk, nor the bulge mass is highly correlated with the galaxies' total ex situ stellar mass.  
The impact of mergers on galaxy structure formation might also depend on the feedback model and could vary across different mass ranges. This topic will be investigated in a forthcoming study (Zeng et al. in prep.).

\begin{acknowledgement}
L.Z. acknowledges the support of the National Key R\&D Program of China No. 2022YFF0503403, and the CAS Project for Young Scientists in Basic Research, Grant No. YSBR-062.   J.C. and X.Y. W. acknowledges the support of the National Natural Science Foundation of China (NFSC) Grant No. 12233005, the National Key Research and Development Program of China (2023YFA1608100) and the China Manned Space Program with grant No. CMS-CSST- 2025-A20.
TNG100 was realized with compute time granted by the Gauss Centre for Super-computing (GCS), under the GCS Large-Scale Project GCS-DWAR (2016; PIs Nelson/Pillepich).

\end{acknowledgement}

\bibliographystyle{aa}

\begin{appendix}

\section{Manually excluded samples}
\label{appendixa}

\begin{figure}[h!]
\centering\includegraphics[width=8cm]{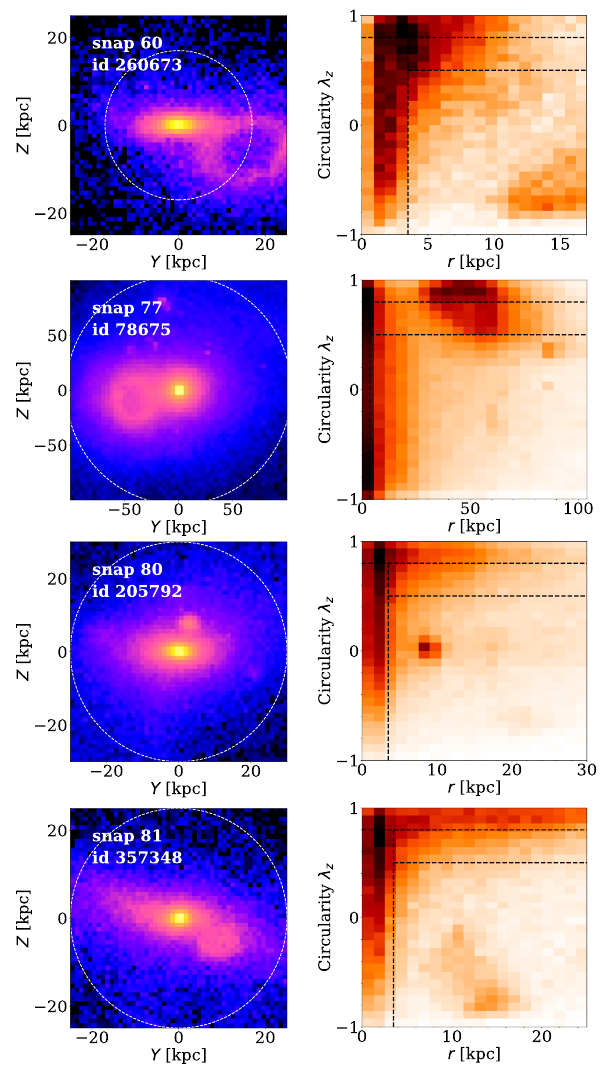}
\caption{Y-Z projection images as well as $\lambda_z$-r phase spaces of representative manually excluded samples mentioned in Section \ref{s:selection}. This figure follows the same plotting convention as Fig. \ref{fig:decomposition}. Both the image and $\lambda_z$-r phase space diagrams reveal clear substructures, indicating these systems are not in dynamical equilibrium. 
}
\label{fig:manually_excluded}
\end{figure}

In Fig. \ref{fig:manually_excluded}, we present the Y-Z image and the corresponding $\lambda_z$ -r phase spaces of four samples that have shown clear substructures and therefore have been manually excluded.

\FloatBarrier

\section{ Dependence of disk survive on \anglemo and \angleso in the range $0^{\circ}-180^{\circ}$}
\label{s:appendixb}

\begin{figure}
\centering\includegraphics[width=8cm]{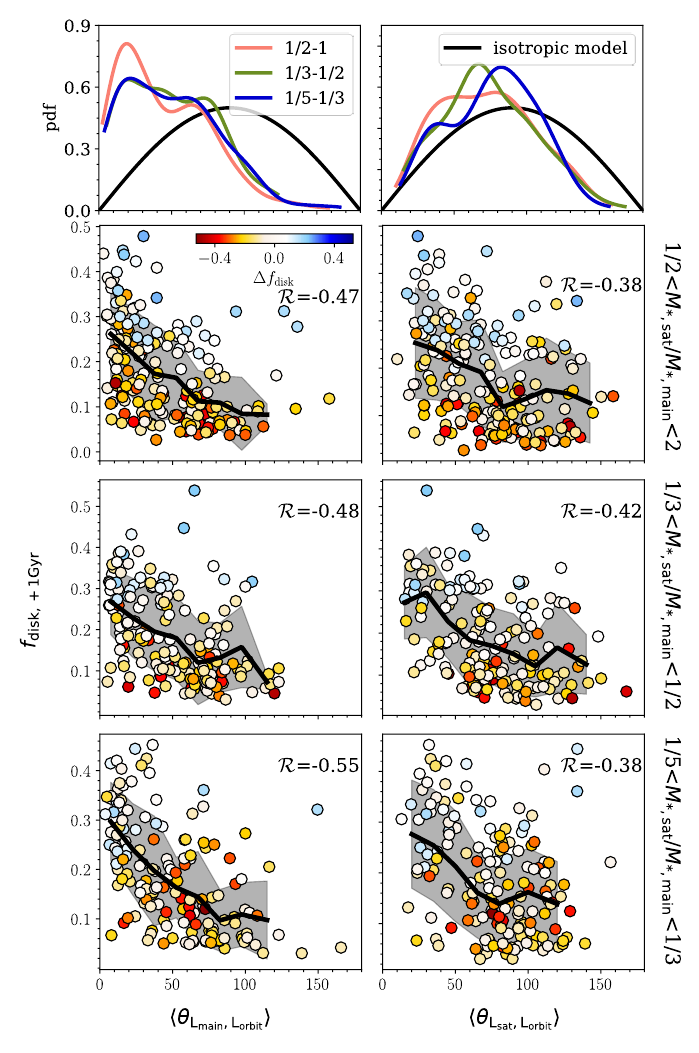}
\caption{Disk fraction in the remnant galaxy at 1Gyr after the merger as a function of \aanglemo and \aangleso. This figure is similar to Fig. \ref{fig:fdiskA}, but the \aanglemo\  and \aangleso\  are not folded.}
\label{fig:appendix}
\end{figure}

In Fig. \ref{fig:appendix}, we further investigate the relationship between disk fraction after merger and \,\aanglemo\ as well as \aangleso, when they are not folded into $0-90^{\circ}$. There are 90\% of our sample have \,\aanglemo\ below $90^{\circ}$, while 98\% have \,\aanglemo\ below $120^{\circ}$. This phenomenon indicates a prevalence of prograde over retrograde orbits, especially for major mergers with mass ratios over 1/2. Moreover, the disk fraction after merger exhibits an overall negative correlation with \aanglemo, with correlation coefficients of -0.47, -0.48, and -0.55 for mass ratios ranging from 1/2 to 2, 1/3 to 1/2, and 1/5 to 1/3, respectively. This trend aligns with Fig. \ref{fig:fdiskA} and can be attributed to the inherent concentration of \,\aanglemo\ values in acute angular regions.

Furthermore, the distribution of \,\aangleso\ closely resembles that of the isotropic model. However, for mergers with mass ratios between 1/5 and 1/2, \,\aangleso\ exhibits a slight inclination towards approximately $90^{\circ}$. This suggests that in mergers with lower mass ratio, the orientation of the satellite galaxies tend to be more perpendicular to the merger orbital plane. Additionally, only 1.5\%, 2.4\%, and 0.6\% of samples exhibit \,\aangleso\ values exceeding 140° for mass ratios between 1/2 to 2, 1/3 to 1/2, and 1/5 to 1/3, respectively, indicating the rarity of the characteristic of satellite galaxies with opposite spin directions to orbital directions.

Interestingly, the disk fraction after merger exhibit a consistent decreasing trend followed by a plateau in relation to \aangleso, regardless of the mass ratio considered, with correlation coefficients of -0.38, -0.42, and -0.38, respectively.

\FloatBarrier

\section{Evolution of the two typical cases}
\label{subsec:typical sample}

\begin{figure*}[h!]
\centering\includegraphics[width=18cm]{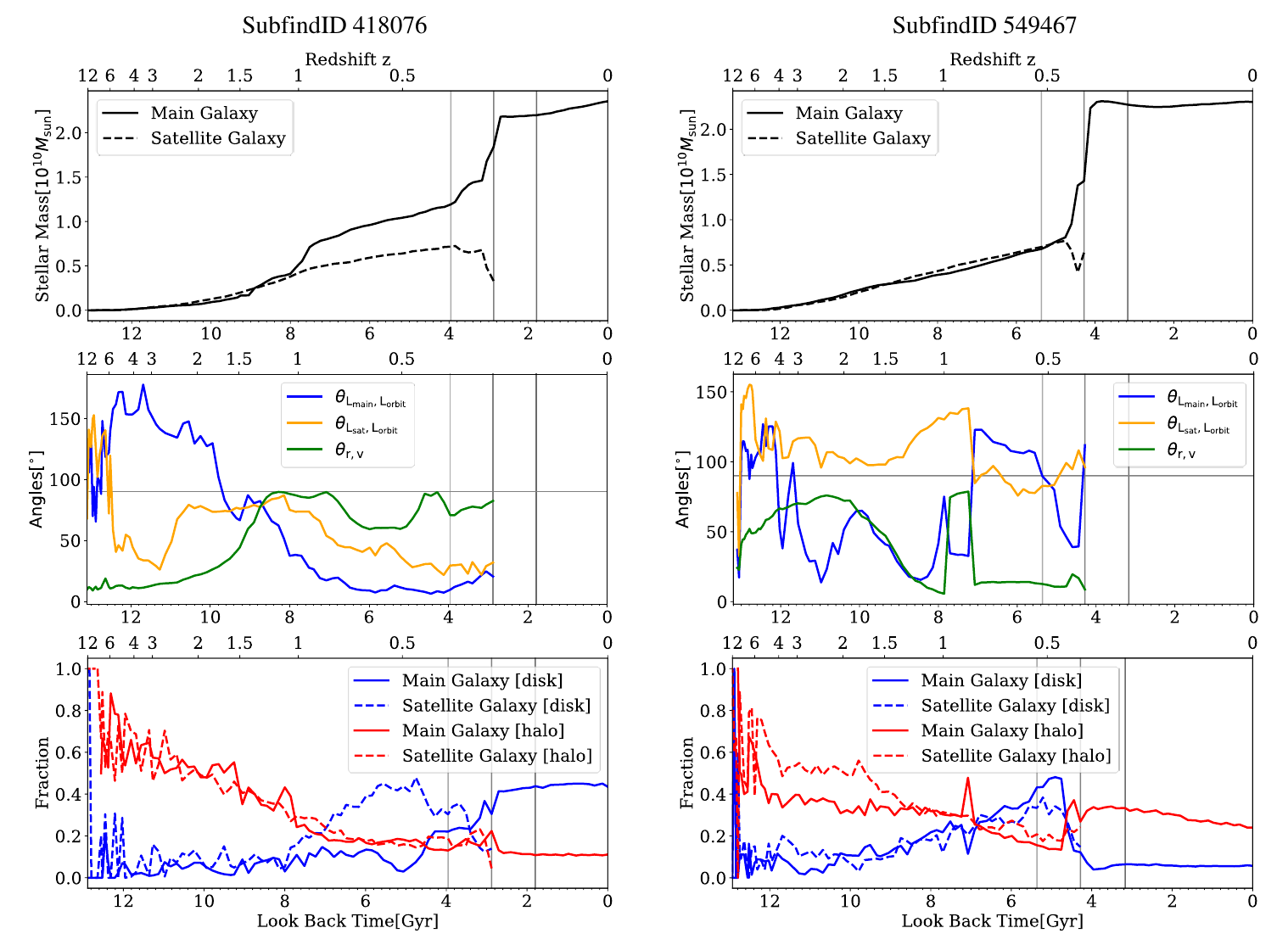}
\caption{Sublink tree of two typical cases in our sample, SubfindID 418076 and 549467 at snapshot 99, respectively. The top two rows of panels display stellar masses and angles over cosmic time, mirroring the format of Fig. \ref{fig:merger_tree} and Fig. \ref{fig:angle_tree}. The bottom row of panels show the fraction of disk and hot inner stellar halo as a function of time. The red and blue lines represent the disk and hot inner stellar halo fractions, while solid lines depict the main galaxy and dashed lines represent the satellite galaxy.
}
\label{fig:merger_tree2}
\end{figure*}

In order to gain a deeper understanding of the two types of scenarios described in Fig. \ref{fig:fdiskAA}, we have selected two representative examples to further examine their merger, as depicted in Fig. \ref{fig:merger_tree2}. In Fig. \ref{fig:fdiskAA} (b), these two examples are marked with blue and red stars, and their corresponding orbital diagrams are shown in panels (c) and (d). The SubfindIDs for these examples at snapshot 99 are 418076 and 549467, respectively.

In Fig. \ref{fig:merger_tree2}, the left column shows the merger corresponding to the orbit depicted in Fig. \ref{fig:fdiskAA} (c). From the second row, it is clear that within 1 Gyr before the merger, the three angles remain stable: \,\anglerv\ stays around $90^{\circ}$, indicating a spiral-in orbit, while \,\anglemo\ and \,\angleso\ fluctuate slightly at a low value, corresponding to a small $\theta_{\rm sum}$. From the third row, we observe that the disk fraction increases significantly after the merger, with $f_{\rm disk, +1 Gyr}$ reaching 0.43. On the other hand, the right column corresponds to the merger of the orbits shown in Fig. \ref{fig:fdiskAA} (d). Here, \,\anglerv\ remains relatively low before merger, suggesting a direct collision orbit. At the same time, \,\anglemo\ exhibits a large range of fluctuations, reflecting the instability of the orbit. In this case, the disk fraction decreases substantially, from 0.43 to 0.06. These two galaxies exemplify the trends shown in Fig. \ref{fig:fdiskAA}: when \,\aanglerv\ is large, $\theta_{\rm sum}$ tends to be small, and $f_{\rm disk, +1 Gyr}$ is high; conversely, when \,\aanglerv\ is small, $\theta_{\rm sum}$ tends to be large, and $f_{\rm disk, +1 Gyr}$ is low.

Additionally, the red lines of the last row illustrate that for the examples shown here, mergers with spiral-in orbits result in a stable or even slightly decreasing fraction of the hot inner stellar halo during the merger. In contrast, mergers with direct-collision orbits lead to a significant increase in the fraction of the hot inner stellar halo. This observation is consistent with the conclusion in Fig. \ref{fig:fdiskAAA}, where it is shown that the changes in the fraction of the hot inner stellar halo vary considerably across different orbits, even when $M_{*,\rm sat}/M_{*,\rm +1 Gyr}$ remains consistent.

\end{appendix}

\end{document}